# Rich Semantic Models and Knowledgebases
# for Highly-Structured Scientific Communication

Robert B. Allen

Yonsei University, Seoul, South Korea

`rballen@yonsei.ac.kr`

**Abstract:** Rather than using text for scientific research reports, we have proposed developing highly-structured reports with rich semantic models. In this paper, we consider detailed structures for the components of research reports using a modeling framework based on a rigorous upper ontology. For instance, we consider the use of structured descriptions of Research Designs to support evaluation of internal and external validity. In addition, collections of highly-structured scientific research reports would be the key component of a set of evolving and interlocking highly-structured scientific knowledgebases.

**Keywords:** Digital Epistemology, Direct Representation, Explanatory Annotations, Internal and External Validity, Research Designs, Scholarly Knowledgebases, Scientific Warrant, Semantic Modeling, Structured Argumentation

## 1 INTRODUCTION

We have proposed the direct representation of scientific research reports with rich semantics and structured applied epistemology [1-10]. In this paper, we consider the requirements of comprehensive highly-structured systems for scholarly communication. As a foundation for the structure, we have been exploring the Basic Formal Ontology (BFO) [11]. We suggest that BFO should be overlaid with additional elements to make it more like a modeling language – perhaps even a form of a programming language [7-9].

Moreover, the results from structured research reports can be used to define or update Reference Ontologies and their associated Models. Cumulatively, research reports from different fields would impact many Ontologies/Models. Thus, the Open Biomedical Ontology (OBO) Foundry [15, 21], which is a collection of specialized ontologies based on the BFO, potentially could be treated as a unified knowledgebase and each of its assertions justified either by common knowledge or by results from the research reports.

Over the last several centuries, scientific research reports have become increasingly structured. Probably the biggest step in that direction was the development of the IMRD (Introduction, Method, Results, Discussion) structure [22, 23]. Several recent projects have proposed additional structure. For instance, ABCDEF [16] proposed flexible linking across the sections of research reports (albeit without full direct representation). Nano-publications and micro-publications have been proposed for describing claims or assertions about scientific research by using Semantic Web structures. Nano-publications [19] are said to be the "smallest unit of publishable information"; they combine an assertion, provenance, and publication information. Micro-publications [14] describe annotations and linking of statements from research reports. In the current version, the emphasis is on natural language statements.

Here, we expand on our earlier proposals for direct representation of research reports and for highly-structured repositories based on those representations. Our top-down approach can be a framework for the further development of the nano and micro-approaches by providing a context for complete research reports, by providing a structured evaluation of the claims, and by linking them into structured scholarly knowledgebases.

## 2  HIGHLY-STRUCTURED RESEARCH REPORTS

### 2.1  Sections of Structured Research Reports

**Introduction:** Well-structured research reports address specific Research Questions. Swales [22, 23] described the function of the Introduction as "Creating a Research Space" (CARS). Thus, the Introduction presents the Research Questions, reviews the relevant literature, and proposes possible answers to the Questions.

Different types of Processes are associated with different types of Research Questions. Some Research Questions might be as simple as whether an effect can be replicated. Others may ask about the generality of an effect (e.g., across species) or about Qualities (e.g., melting points). If the Research Questions relate to a causal mechanism, then the Research Space would be a list of plausible causal mechanisms that are consistent with existing models. Those tentative models could be specified to varying levels of detail with semantic structures such as described by [9], with placeholder entities filling any gaps. In testing for causal processes, a typical strategy would be to study the impact of disrupting one of the components. If the disruption affects the target, that is evidence that the disrupted component and the path with which it is associated play a role in the target process. The interpretation of the results depends heavily on the selection of plausible models available from the knowledgebases.

The Introduction should also present the motivation for the Research Questions. It is challenging with the current generation of modeling tools to provide structured descriptions for motivations because motivations are often general, such as advancing knowledge or fulfilling human needs. However, frameworks may ultimately be developed, for instance, by grounding some human needs in biology.

**Research Methods and Designs:** Research Methods are workflows. [12] recognizes the importance of research workflows and discusses how ad hoc research workflows can be constructed from existing ontology elements. However, this seems ad hoc and does not include unified workflow structures. To provide a workflow framework that meshes well with other elements, we propose specific workflow structures as part of the Model Layer [9].

In addition, we propose that Research Methods should be modeled in the context of Research Designs (e.g., experimental and quasi-experimental designs [13]). Research Designs can provide a systematic framework for linking the Research Space, Research Methods, and the Results.

**Results:** Semantic representations may be used to describe research data (see [20] for an example). Such representations should be systematically collected and linked back to the Research Space and Research Methods. Moreover, structured Research Designs can provide guidelines for comparing results from different conditions [3, 6] and ultimately support alternatives models in the Research Space.

Beyond simply describing the data, the Research Report must evaluate and make claims about of validity [13]. Internal validity refers to whether the Research Methods were applied successfully. In combination with the Research Methods and Results, details about Techniques and Instrumentation[1] can address threats to Internal Validity. "Checks on the manipulation" confirm whether the manipulation worked as intended.

External validity refers to the extent to which the Research Results can be generalized to other situations. The expectations about generalization are derived from the models. Thus, External Validity depends on the extent to which the results address the Research Questions and whether the results are compatible with the Reference Ontologies/Models. In a few cases, research may show unexpected results and lead to restructuring the Ontologies/Models.

**Discussion:** The Discussion in a research report summarizes the implications of Results for the Research Questions. It may propose revisions to the Reference Ontologies/Models based on those claims. The Discussion may also comment on implications of the Results for research issues beyond the Research Questions.

### 2.2   Structured Research Reports as Supporting Structured Applied Epistemology

The nature of scientific research reports as a type of discourse has long been noted. In our previous work, we highlighted the relevance of work on discourse for highly-structured scientific research reports. In [8], we examined the relationship of claims to making updates of the knowledgebases (i.e., epistemology) (see also [24]). Structured scientific research reports are a particularly rich type of structured argumentation [4, 25]. For example, the evaluation of internal and external validity may be viewed as addressing the Premises and Critical Questions [25].of the Research Design.

Claims-making is related to speech acts. There is an extensive literature on "speech acts" associated with argumentation generally, but only [17] examines claims made in scientific research reports. Scientific research report speech acts cover the range of logical outcomes for the research based on plausible models for it. Moreover, research claims made in published research reports may be considered as documenting a commitment (i.e., a document act) which asserts the validity of the research findings.

## 3   HIGHLY-STRUCTURED SCHOLARLY KNOWLEDGEBASES

### 3.1   Interlocking Knowledgebases

Individual research reports as described in Section 2 need to be collected and then coordinated with other repositories. Here we describe some of the requirements and challenges for such structured repositories.

- **Research Report Repository:** Presumably, each structured research report would be accompanied by a standard text version much like a traditional research report. However, that standard text version might not be written in the traditional sense; rather, it could be generated from the knowledge-structures using repository the management tools (see below).

  In a structured report repository, there could be a variety of links between the elements. Rather than traditional citations, topic links between relevant sections of the research reports could be implemented [6, 16].  The authors of the target article would be known but the topic and not the author would be the focus of the links.

---

[1]  In PLOS, "Techniques and Instrumentation" is called "Materials and Methods" and it appears at the end of the articles.

- **Reference Ontologies/Models:** The Reference Ontologies/Models would include and extend existing ontologies, such as those in the OBO Foundry.  The claims about updates to the Reference Ontologies/Models would be justified through links to the research reports and supported with the Discussion; updates may eventually be implemented once the claims gain the consensus of the research community.

  While several important policies have been established for OBO [15], much tighter specifications will be required to create rich semantic knowledgebases.  More than simple semantic networks, BFO-based ontologies have some features of classification systems because they use a single-threaded is_a hierarchy.  The upper ontology provides a sort of faceting.  Moreover, as a realist, science-oriented ontology BFO may employ scientific warrant.

- **Tentative Models/Ontologies:** The Reference Ontology/Models are based on consensus acceptance of results but there are many open Research Questions about which there is no consensus.  We propose a sandbox for exploring models that may have considerable support but do not reach consensus.  These tentative models should be open to rigorous commentary and debate.  Because there is no absolute measure of consensus, the tentative models will blend into the Reference Ontologies/Models as described in the previous section.

- **Research Designs, Standard Methods, Techniques, and Instrumentation:** There could be a collection of standard workflow templates, details about instruments, and manuals of best practices.  Moreover, there could be links to research papers that focus on the development of new Research Methods rather than specific Research Questions.  Some recent work in digital preservation [18] describes developing collections of scientific research workflows.  Potentially those collections could be extended with rich semantic structures and linking to structured Results and Research Questions.

- **Reference Datasets:** Standard reference datasets such as collections of physical constants (e.g., melting points) and catalogs (e.g., of astronomical objects, geographic features) should be included.  Some of these reference values would be based on research reports and links to those sources should be noted.  Detailed data should be collected from all research reports; those data would most naturally be associated directly with the research reports.

- **Authority Records and Structured Scholarly Editions:** Collections of bibliographic data include "authority records".  These are verified records about authors and corporate entities associated with publications.  The ORCID project provides linked data about many authors of research reports.  These are a natural foundation for structured authority records.  Other projects are collecting rich biographical data for archival records; at least in some cases, authority records and rich biographical records could be coordinated.  Even more broadly, the entire publication history could be integrated into a "unified temporal map" such as proposed for structured scholarly editions [10].  This would provide a "provenance view" of the publication history rather than simple metadata tags.

- **Case Studies:** Models of Particulars should be able to describe Scenarios.  However, in many situations (e.g., medical diagnosis) the descriptions may be complex and a detailed case study is needed.  Such case studies could be richly structured and linked into other knowledgebases.

- **Secondary Literature and Explanatory Annotations:** In addition to primary research repositories, there could be a large Secondary Literature. This could include materials such as surveys, textbooks, magazine articles, visualizations, and videos, or even virtual reality documentaries. Because they would be linked to Research Reports and other knowledgebases and because they might clarify details for different audiences, we might call them explanatory annotations. Some of these secondary materials might be stored as text but they could also be derived from loose scripts and hypertext guided tours (see [3, 5]) that could be personalized to generate different versions based on user knowledge and interests (see User Tools section below).

Many types of Claims are made in research reports and related materials. For instance, some claims are about the implications of research results while others are about internal and external validity. The breakdown here could be considered an initial taxonomy of Claim types.

### 3.2 Services for Knowledgebases

The repositories should also support services such as:

- **Versions and Maintenance:** Because knowledgebases are evolving, version management will be needed to support updates while also providing access to previous configurations.

- **Evaluation and Review:** Editors and reviewers would provide comments and overall evaluation for the proposed knowledgebases. Tools should be developed to support the comments and for later inspection and summarization. Beyond the tools, a social infrastructure is needed to establish policies for updates to the knowledgebases.

- **User Tools:** Some tools would support text generation for structured descriptions. They could perform text generation across languages and might go beyond text to develop multimedia and multimodal presentations. Beyond simply generating explanations, the tools could provide tutorials (see [3, 5] Sections 10.2, 10.3).

## 4  CONCLUSION

We have provided an overview of the development of highly-structured research reports and scholarly repositories. We propose using models based on an extended version of the Basic Formal Ontology and to develop structured versions of the Introduction, Method, Results, and Discussion of research reports. We emphasize the central roles of modeling the Research Design and of the evaluation of Internal and External Validity. Moreover, we consider how these research reports can be organized into a broader infrastructure of interlocking scientific knowledgebases.